\documentclass[10pt,letterpaper,twocolumn]{article} 

\usepackage{ol2}
\usepackage[draft]{hyperref}
\usepackage{amsmath}

\begin{document}

\twocolumn[ 

\title{ Light propagation in a birefringent plate with topological
charge}

\author{Ebrahim Karimi,$^{1,2}$ Bruno Piccirillo,$^{1}$ Lorenzo
Marrucci,$^{1,2}$ and Enrico Santamato$^{1,*}$}

\address{$^{1}$Dipartimento di Scienze Fisiche, Universit\`{a}
degli Studi di Napoli ``Federico II'', Complesso di Monte S. Angelo,\\
via Cintia, 80126 Napoli, Italy\\
$^{2}$Consiglio Nazionale delle Ricerche-INFM Coherentia, Napoli,
Italy\\$^*$Corresponding author: enrico.santamato@na.infn.it }

\begin{abstract}
We calculated the Fresnel paraxial propagator in a birefringent
plate having topological charge $q$ at its center, named
``$q$-plate''. We studied the change of the beam transverse profile
when it traverses the plate. An analytical closed form of the beam
profile propagating in the ``$q$-plate'' can be found for many
important specific input beam profiles. We paid particular attention
to the plate having a topological unit charge and we found that if
small losses due to reflection, absorption and scattering are
neglected, the plate can convert the photon spin into orbital
angular momentum with up to 100~\% efficiency, provided the
thickness of the plate is less than the Rayleigh range of the
incident beam.
\end{abstract}


 ] 

\begin{figure}[h]
\centerline{\includegraphics[width=7.5cm]{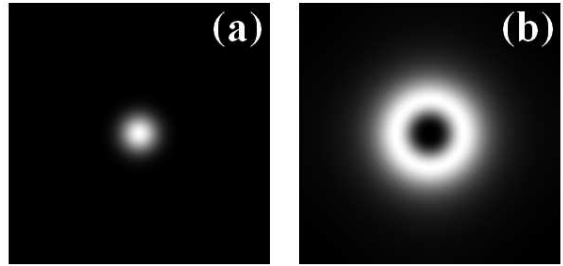}}
\caption{\label{fig:3}  Intensity profile in the far-field beyond
the 1-plate after free-air propagation. (a) No STOC; (b) Full STOC.}
\end{figure}

\noindent Light beams carrying orbital angular momentum (OAM) are
receiving increasing attention as a resource in quantum and
classical optics, since OAM exists in an inherently multidimensional
space. Information can thus be encoded in higher dimensional
OAM-alphabets~\cite{molinaterriza02,molinaterriza07} for its use in
free-space communication systems~\cite{gibson04} or to increase the
dimensionality of the working Hilbert space in quantum
communications systems~\cite{mair01}. The main characteristics of a
light beam carrying OAM is the presence of a topological
point-charge of integer order in the optical phase. It was recently
demonstrated that a birefringent plate made of a vortex-patterned
liquid crystal film can imprint its topological charge into the
optical phase of the incident light, thus changing the OAM of the
beam. Being birefringent, the plate affects the photon spin angular
momentum (SAM) too, thus providing easy control of the beam OAM
content by either changing its polarization or changing the
retardation of the plate~\cite{marrucci06}. It can be shown that, if
the topological charge of the birefringent plate is $q$, the OAM of
a light beam passing through such a ``$q$-plate'' (QP), changes by
an amount ${\ensuremath\pm}2q\hbar$ per photon. One may roughly
think of the QP as of an ordinary birefringent plate rotated at
angle $\alpha$ about the beam $z$-axis, with $\alpha$ given by
$\alpha=\alpha(x,y)=\arctan(y/x)=\phi$ where $\phi$ is the azimuthal
angle in the $x,y$-plane. This simple picture is enough to account
for some qualitative effects such as, for instance, the SAM-to-OAM
conversion (STOC)~\cite{marrucci06}, the associated optical Berry
phase~\cite{marrucci06a} and the changes of the SAM and OAM content
of the beam at different depths in the QP~\cite{calvo07}, but cannot
be used to determine the detailed quantitative behavior of the beam
during its propagation. In fact, the formal replacement
$\alpha\rightarrow\phi(x,y)$ is justified only if the function
$\alpha(x,y)$ varies smoothly over the optical wavelength scale (the
so-called geometric optics approximation (GOA))~\cite{calvo07},
which is not true in the present case, because no length scale is
defined by the function $\arctan(y/x)$. As shown below, exact series
solution of the wave equation inside the QP shows that all field and
wave propagators at the QP singularity vanish which is not true for
fields \cite{marrucci06} and propagators \cite{calvo07} calculated
by just setting $\alpha\rightarrow\phi$.
To overcome this broad discrepancy, in this work, we study the
propagation of a light beam in the QP without having recourse to the
GOA, but assuming, instead, a good beam paraxiality. We start from
Maxwell's wave equation $\nabla^2\mathbf
E-\nabla(\nabla{\ensuremath\cdot}\mathbf E)+
k_0^2\,\hat{\epsilon}{\ensuremath\cdot}\mathbf E=0$, where
$k_0=2\pi/\lambda=\omega/c$, $\hat\epsilon$ is the relative
dielectric tensor at frequency $\omega$, $c$ is the speed of light
and  $\lambda$ is the wavelength. The relative dielectric tensor is
given by
$\hat\epsilon=\hat{R}(q\phi)\cdotp\hat\epsilon_{\textrm{local}}\cdotp\hat{R}(-q\phi)$,
where $\hat{\epsilon}_{\textrm{local}}={\rm
diag}(n_o^2,n_e^2,n_o^2)$ is the dielectric tensor in the local
frame of the plate, $\hat{R}(\alpha)$ is the rotation matrix about
the $z$-axis, $\phi$ is the azimuthal angle, $n_o$ and $n_e$ are the
material ordinary and extraordinary indices, respectively. We
neglect absorption and assume positive birefringence $n_e>n_o$. In
most materials, including liquid crystals, the birefringence is
small $(n_e-n_o)\ll n_o$, so that we may neglect the longitudinal
part of the optical field and take $\nabla{\ensuremath\cdot}\mathbf
E\simeq 0$. In this approximation, the wave equation reduces to
Helmholtz's vector equation $\nabla^2\mathbf
E_\perp+k_0^2\hat\epsilon{\ensuremath\cdot}\mathbf E_\perp=0$ for
the transverse part $\mathbf E_\perp$ of the field. In view of the
cylindrical symmetry of the problem, it is convenient to find the
eigenmodes of the Helmholtz's vector equation in the circular
polarization basis $E_\pm = (E_x\pm i E_y)/\sqrt{2}$ and in the
cylindrical coordinates $(r,\phi,z)$, by setting $\mathbf
E_\perp(r,\phi,z)=({\cal E}_+(r)e^{i(m+q)\phi},{\cal
E}_-(r)e^{i(m-q)\phi},0)e^{-ik_0\gamma z+i\omega t}$, where $\gamma$
is the longitudinal spatial frequency and $z=0$ is the input-face of
the QP. Inserting this field into Helmholtz's equation, yields a
pair of coupled radial equations
\begin{eqnarray}\label{eq:fgODE}
   f''(r)+\frac{f'(r)}{r}+\left(k_0^2(n_o^2-\gamma^2)-\frac{\mu^2}{r^2}\right)f(r)=\frac{\nu\,g(r)}{r^2}
  \cr
   g''(r)+\frac{g'(r)}{r}+\left(k_0^2(n_e^2-\gamma^2)-\frac{\mu^2}{r^2}\right)g(r)=\frac{\nu\,f(r)}{r^2}
\end{eqnarray}
where $f(r)=({\cal E}_+ + {\cal E}_-)/\sqrt{2}$, $g(r)=({\cal E}_+ -
{\cal E}_-)/\sqrt{2}$, $\mu=\sqrt{m^2+q^2}$ and $\nu=2m\,q$ .
Equations (\ref{eq:fgODE}) are exact and can be solved by series. We
postpone a full discussion about their solutions to a future work.
Here we consider only the approximate solutions for paraxial beams
at normal incidence. Setting
$\gamma=\gamma_o=\sqrt{n_o^2-\beta^2}\simeq n_o-\beta^2/2n_o$, where
$\beta$ is a transverse spatial spectrum, Eqs.(\ref{eq:fgODE})
reduce to the equations for the ordinary wave
\begin{eqnarray}\label{eq:fgOwave}
   f_o''(r)+\frac{f_o'(r)}{r}+\left(k_0^2\beta^2-\frac{\mu^2}{r^2}\right)
   f_o(r)=\frac{\nu\,g_o(r)}{r^2}
   \cr
   g_o''(r)+\frac{g_o'(r)}{r}+\left(k_0^2\beta^2\Lambda^2-\frac{\mu^2}{r^2}\right)
   g_o(r)=\frac{\nu\,f_o(r)}{r^2}
\end{eqnarray}
where $\Lambda^2=1+2(n_e^2-n_o^2)/\beta^2$. We observe that in
commercial liquid crystals we have $n_e^2-n_o^2\simeq 0.5$ while
usual paraxial laser beams at normal incidence have the radial
spatial frequency with $\beta$ ranging from zero to $\beta\simeq
10^{-2}$. The parameter $\Lambda^2$ is therefore very large in all
practical cases. We may then solve Eqs.(\ref{eq:fgOwave}) as
asymptotic series of $\Lambda^2$. The zero-order approximation of
the asymptotic solution of Eq.(\ref{eq:fgOwave}) for the ordinary
wave is given by $f_o(r) = A_o J_\mu(k_0 \beta r)$ and $g_o(r)=0$,
where $A_o$ is an arbitrary constant and $J_\mu(x)$ is
Bessel'function of index $\mu={\sqrt{m^2+q^2}}$. The differential
equation for the extraordinary wave is obtained from
Eqs.(\ref{eq:fgODE}) by setting
$\gamma=\gamma_e=\sqrt{n_e^2-\beta^2}\simeq n_e-\beta^2/2n_e$ and it
can be obtained from Eqs.(\ref{eq:fgOwave}) by the formal
replacements $f\rightarrow g,g\rightarrow f,\Lambda^2\rightarrow
-\Lambda^2$. The zero-order asymptotic solution for the
extraordinary wave is then given by $f_e(r)=0$ and $g_e(r) = A_e
J_\mu(k_0 \beta r)$ with constant $A_e$. All other terms of the
asymptotic solution can be found recursively. The zero-order
asymptotic solutions hold in the whole $x,y$-plane except a small
region about the origin having a radius
$r_0\approx\lambda/(\Lambda\beta)=\lambda/\sqrt{n_e^2-n_o^2}$. For
commercial liquid crystal we have $r_0\approx 1.25\lambda$. In this
small region is located the singularity and the optical axis is not
well defined here. The effect of this region can be accounted for
only by the exact (not paraxial) wave approach, but we may
anticipate on physical grounds that the main effect of this region
is to scatter a small fraction of the light at large angles out of
the paraxial beam. From the asymptotic paraxial modes of Helmholtz'
equation it is straightforward to calculate the Fresnel paraxial
propagator for the optical field $\mathbf E_\perp$. The optical
field at plane $z$ in the QP is given by
\begin{eqnarray}\label{eq:Eperp}
    \mathbf E_\perp(r,\phi,z)&=&\frac{1}{2}\int_0^\infty
    \!\rho\,d\rho\int_0^{2\pi}\!d\psi\,\hat R(q\phi)
        {\Big\lbrace} \left({K^o+K^e}\right)\hat 1\cr
    &+&\left({K^o-K^e}\right)\hat\sigma_z{\Big\rbrace}
         \hat R(-q\psi)\mathbf E_\perp(\rho,\psi,0).
\end{eqnarray}
where $\hat 1$, $\hat R(\phi)$ and $\hat\sigma_z$ are the
$2{\ensuremath\times} 2$ unit, rotation and Pauli's matrices,
respectively. The Fresnel kernels in Eq.(\ref{eq:Eperp}) are given
by $K^{o,e}=\sum_{m}
    K^{o,e}_{\mu(m)}(r,\rho;z)e^{im(\phi-\psi)}$, where $\mu(m)=\sqrt{m^2+q^2}$ and
\begin{eqnarray}\label{eq:KoKe}
  K^{o,e}_{\mu(m)}(r,\rho;z) &=& \left(\frac{i n_{o,e} k_0}{2\pi z}\right)
    i^{\mu(m)} J_{\mu(m)}\left(\frac{k_0 n_{o,e}r\rho}{z}\right)\cr
    &\times& e^{-\frac{ik_0 n_{o,e}(r^2+\rho^2)}{2z}-ik_0n_{o,e} z}
\end{eqnarray}
The Fresnel kernels $K^o$ and $K^e$ in Eq.(\ref{eq:KoKe}) are
characterized by the presence of Bessel function of irrational
order. Although, $K^o$ and $K^e$ cannot be obtained in a closed
form, they permit to evaluate analytically the field transmitted by
the QP in important cases as, for instance, for Laguerre-Gauss incident beams.
Here we consider only the case of a LG$_{0l}$ beam
impinging onto the QP. Setting $\mathbf
E_\perp(\rho,\phi,0)=e^{il\phi}\hbox{LG}_{0}^{l}(\rho)\left[\begin{array}{c}
a \\ b \\ \end{array}\right]$ in the circular polarization basis,
where $\hbox{LG}_{0}^{l}(\rho)$ is the radial amplitude of
Laguerre-Gauss modes, we obtain
\begin{equation}\label{eq:LG0}
    \left[\begin{array}{c} E_+ \cr
    E_- \\ \end{array}\right]
    = e^{i(l\phi-k_0 n_o z)}
    \left[\begin{array}{cc} K^{+}_{\mu^{-}} & K^{-}_{\mu^{+}} e^{2iq\phi} \cr
    K^{-}_{\mu^{-}} e^{-2iq\phi} & K^{+}_{\mu^{+}}  \\
    \end{array}\right]
    \left[\begin{array}{c} a \cr
    b \\ \end{array}\right]
\end{equation}
where $K^{\pm}_{\mu}=(\hbox{HyGG}_{|l|-\mu,\mu}(r,z/n_o)\pm
e^{-ik_0\Delta n\,z}\hbox{HyGG}_{|l|-\mu,\mu}(r,z/n_e))/2$, $\Delta
n=n_e-n_o$, $\mu^{\pm}=\mu{(\ell\pm q)}$ and
$\hbox{HyGG}_{p,m}(r,z)$ is the Hypergeometric-Gaussian
mode~\cite{karimi07}, viz.
\begin{eqnarray}\label{eq:HyGG}
   \hbox{HyGG}_{pm}(\rho,\zeta)&=&C_{pm}\,\zeta^{\frac{p}{2}}(\zeta+i)^{-(1+|m|+\frac{p}{2})} \rho^{|m|}\\
    &{\ensuremath\times}&\,e^{-\frac{i\rho^2}{(\zeta+i)}}
   {}_1\!F_1\left(-\frac{p}{2},1+|m|;\frac{\rho^2}{\zeta(\zeta+i)}\right)\nonumber
\end{eqnarray}
where $C_{pm}=i^{|m|+1}\sqrt{\frac{2^{p+|m|+1}}{\pi\Gamma(p+|m|+1)}}
    \frac{\Gamma\left(1+|m|+\frac{p}{2}\right)}{\Gamma\left(|m|+1\right)}$,
    $\rho=r/w_0$, $\zeta=z/z_R$ and $z_R=k_0w_0^2/2$ is the
beam Rayleigh range. Because $n_o\simeq n_e$, the arguments of the
function $\hbox{HyGG}_{pm}$ in Eq.(\ref{eq:LG0}) are very close, so
that when $\Delta n\,z=j\lambda$ ($j=1,2,\dots$) the matrix in
Eq.(\ref{eq:LG0}) is almost diagonal, the beam in the QP has the
same value of OAM, i.e. $\ell\hbar$ per photon. When $\Delta n \,z =
(2j-1)\lambda/2$, instead, only the off-diagonal elements survive,
the right and left circular components of transmitted field assume a
phase factor $e^{{\ensuremath\pm}2iq\phi}$ and the beam OAM change
by ${\ensuremath\pm}2q\hbar$ per photon, depending on the input
circular polarization helicity. As the beam propagates in the QP,
its transverse profile, spin and OAM change. From Eq.(\ref{eq:LG0}),
we may calculate the average SAM and OAM carried by the beam at the
plane $z$ in the QP, obtaining
\begin{align}\label{eq:sz}
   S_z(z)=\frac{1}{\omega}\Re\big[e^{-ik_0\Delta nz}\big(&|b|^2\,
   I_{|\ell|-\mu^+,\mu^+}(z)\cr &-|a|^2\,I_{|\ell|-\mu^-,\mu^-}(z)\big)\big]
\end{align}
\begin{equation}\label{eq:Lz}
  L_z(z)+\frac{q}{\omega}S_{z}(z)=\frac{1}{\omega}\left(\left(\ell-q\right)
  |a|^2+\left(\ell+q\right)|b|^2\right)
\end{equation}
where
\begin{eqnarray}\label{eq:Ip}
   I_{p,m}(\zeta) &=&
   \frac{2^{p+|m|+1}\Gamma^2\left(\frac{p}{2}+|m|+1\right)}{\Gamma\left(|m|+1\right)
   \Gamma\left(p+|m|+1\right)}\,\chi^{-p/2}(\zeta)\cr
          \;& \times&\left(\frac{n_o\,n_e}{2n_o\,n_e-i(n_e-n_o)\zeta}\right)^{p+|m|+1}
          \cr\;& \times&\,{}_2\!F_1\left(-\frac{p}{2},-\frac{p}{2};|m|+1;\chi(\zeta)\right)
\end{eqnarray}
and  $\chi(\zeta) = \left(\frac{n_e n_o}{n_e
n_o-i(n_e-n_o)\zeta}\right)^2$. As expected, we have $I_{p,m}(0)=1$
so that Eqs.(\ref{eq:Lz}) and (\ref{eq:sz}) yield
$S_z(0)=(|b|^2-|a|^2)/\omega$ and $L_z(0)=(|a|^2+|b|^2)\ell/\omega$.
\begin{figure}[htb]
\centerline{\includegraphics[width=9cm]{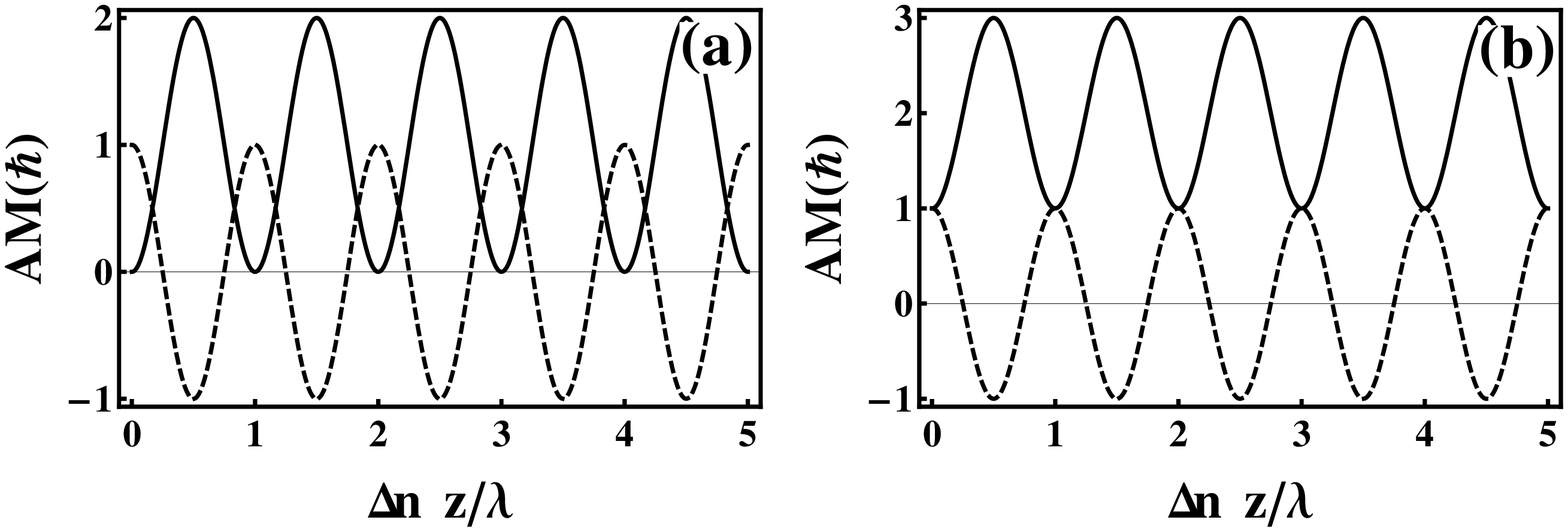}}
\caption{\label{fig:1} Beating of SAM (dashed line) and OAM (full
line) as a function of the optical retardation $\Delta n z/\lambda$
while a circularly polarized input beam propagates in the 1-plate.
(a) For LG$_{00}$ and (b) LG$_{01}$ as a input beam.  We used the
following data: $n_o=1.5$, $n_e=1.7$, $w_0=50\lambda$.}
\end{figure}
In Fig.(\ref{fig:1}) the photon STOC~\cite{marrucci06} is shown as a
function of the propagation depth in the 1-plate for LG$_{00}$ and
LG$_{01}$ input beams. The conversion efficiency is practically
100\% and its maximum occurs at optical retardation $\Delta
n\,z=(2j-1)\lambda/2$ with integer $j$. When the optical retardation
is $j\lambda$, no conversion occurs and the beam has no OAM.
Changing the optical retardation of the 1-plate provides a good way
to control the STOC process. However, when the thickness of the
1-plate becomes very large (much larger than the beam Rayleigh
range) the conversion efficiency slowly decays.
\begin{figure}[htb]
\centerline{\includegraphics[width=9cm]{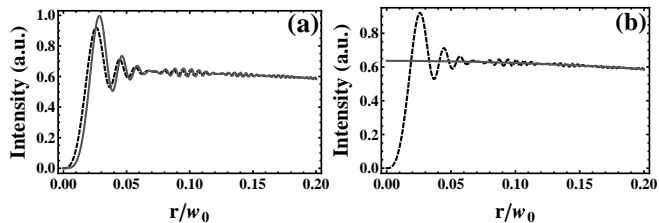}}
\caption{\label{fig:2} Intensity profile for (a) full STOC (b) no
STOC in the 1-plate. Full and dashed lines are simulated by
\cite{calvo07} and our theory, respectively. The input beam assumed
the TEM$_{00}$.}
\end{figure}
According to Eqs.(\ref{eq:LG0}) and (\ref{eq:HyGG}), the field
profile inside the 1-plate (and at its exit face) vanishes as
$r^{\sqrt{2}}$ along the beam axis so that the intensity profile has
the characteristic doughnut shape irrespective of the OAM carried by
the beam. Fig. \ref{fig:2} shows the intensity profiles for (a) full
STOC (b) no STOC. For the sake of comparison, the results obtained
in the GOA~\cite{calvo07} are also shown. We can deduce that the GOA
approximation is fairly good for the case of full STOC, but is very
bad in the near field and in the case of no STOC. Dramatic changes
of the intensity profile depending on the final OAM are seen,
however, in the far-field after free-air propagation. When the STOC
is maximum, in fact, we observe the doughnut profile, while when no
conversion occurs, the far-field pattern has again a maximum at its
center. This is shown in Fig. (\ref{fig:3}).\newline
In conclusion, we calculated the Fresnel propagator of an optical
beam in a birefringent plate having a topological charge $q$ in the
paraxial approximation and for normal incidence. We considered in
the some details the propagation of a LG$_{0\ell}$ beam in the QP.
As the beam traverses the plate, its transverse profile changes from
Laugerre-Gaussian to Hypergeometric-Gaussian and STOC occurs. We
paid particular attention to the 1-plate and we found that the
conversion efficiency is almost 100\% when the thickness of the
plate is much smaller than the beam Rayleigh range and slowly
decreases when the thickness of the 1-plate is increased. The free
propagation of the HyGG modes generated by the QP is not stable and
the characteristic transition from dot-to doughnut profile when the
OAM changes from 0 is observed only in the far field. The
possibility offered by the azimuthally oriented plate in
manipulating entanglement among several degrees of freedom of the
light may be of great interest for quantum information, quantum
communications and quantum computing.


\begin{thebibliography}{99}

\bibitem{molinaterriza02}
G.~Molina-Terriza, J.~P. Torres, and L.~Torner, \prl \textbf{88},
013601 (2002).

\bibitem{molinaterriza07}
G.~Molina-Terriza, J.~P. Torres, and L.~Torner, Nat.\ Phys.
\textbf{3}, 305  (2007).

\bibitem{gibson04}
G.~Gibson, J.~Courtial, M.~J. Padgett, M.~Vasnetsov, V.~Pasko, S.~M.
Barnett,  and S.~Franke-Arnold, \opex \textbf{12}, 5448 (2004).

\bibitem{mair01}
A.~Mair, A.~Vaziri, G.~Welhs, and A.~Zeilinger, \nat \textbf{412},
313 (2001).

\bibitem{marrucci06}
L.~Marrucci, C.~Manzo, and D.~Paparo, \prl \textbf{96}, 163905
(2006).

\bibitem{marrucci06a}
L.~Marrucci, C.~Manzo, and D.~Paparo, \apl \textbf{88}, 221102
(2006).

\bibitem{calvo07}
G.~F. Calvo and A.~Pic{\'{o}}n, \ol \textbf{32}, 838 (2007).

\bibitem{karimi07}
E.~Karimi, G.~Zito, B.~Piccirillo, L.~Marrucci, and E.~Santamato,
\ol \textbf{32}, 3053 (2007).
\end{thebibliography}
\end{document}